# Energy loss of terahertz electromagnetic waves by nano-sized connections in near-self-complementary metallic checkerboard patterns

**Keisuke Takano,[1,2*] Yoku Tanaka,[1] Gabriel Moreno,[3] Abdallah Chahadih[3], Abbas Ghaddar[3], Xiang-Lei Han[3], François Vaurette[3], Yosuke Nakata[2,4], Fumiaki Miyamaru[2,5], Makoto Nakajima[1], Masanori Hangyo[1,**], and Tahsin Akalin[3]**

[1]*Institute of Laser Engineering, Osaka University, 2-6 Yamadaoka, Suita, Osaka 565-0871, Japan*

[2]*Center for Energy and Environmental Science, Shinshu University, 4-17-1 Wakasato, Nagano 380-8553, Japan*

[3]*Institute of Electronics, Microelectronics and Nanotechnology (IEMN), Lille 1 University, Villeneuve d'AscqCedex, France*

[4]*Research Center for Advanced Science and Technology, The University of Tokyo, 4-6-1 Komaba, Meguro-ku, Tokyo 153-8904, Japan*

[5]*Department of Physics, Faculty of Science, Shinshu University, 3-1-1 Asahi, Matsumoto, Nagano 390-8621, Japan*

*Corresponding: ksk_takano@shinshu-u.ac.jp,

**Deceased (October 25, 2014)

**Abstract:** The design of a self-complementary metallic checkerboard pattern achieves broadband, dispersion-less, and maximized absorption, concentrating in the deep subwavelength resistive connections between squares, without any theoretical limitation on the energy absorbing area. Here, we experimentally and numerically investigate the electromagnetic response in the limit of extremely small connections. We show that finite conductivity and randomness in a near-self-complementary checkerboard pattern plays a crucial role in producing a frequency-independent energy loss in the terahertz frequency region. Here metals behave like an almost perfect conductor. When the checkerboard pattern approaches the perfect self-complementary pattern, the perfect conductor approximation spontaneously breaks down, owing to the finite conductivity at the nano-scale connection, leading to broadband absorption. It is also shown that the random connections between metallic squares also lead to broadband and maximized energy loss through scattering loss, similar to finite conductivity.

**I.     Introduction**

The miniaturization of the light receiving area in photonic or electromagnetic devices improves the efficiencies and operating speeds, but prevents coupling with incident electromagnetic waves. The electromagnetic resonances of antennas, frequency selective surfaces,[1] metasurfaces (2-dimensional metamaterials[2]), and plasmonic structures[3], can be used to improve the coupling. Although the current concentration caused by electromagnetic resonance in a device improves the coupling in the small light receiving area, this improvement is limited by a narrow frequency band. In contrast, a self-complementary design[4] satisfies both the broadband operation and the highly efficient coupling between the deep sub-wavelength structures and the electromagnetic waves.

Recently, Urade et al. demonstrated that the broadband energy absorption for terahertz waves in a self-complementary checkerboard pattern [Fig. 1(a)] consisted of aluminum squares terminated with micro-meter sized titanium films.[5] The broadband absorption of terahertz waves is concentrated into the titanium resistive sheets. In their paper, the energy absorption area was chosen to have a similar scale of metallic squares, but there was no theoretical limitation on the size of the resistive sheets. Therefore, the broadband efficient energy transfer into an extremely small scale can be produced by shrinking the resistive elements infinitely. This is desirable in the design of devices such as detectors, sensors, and modulators.[6–8] At terahertz frequencies, this is especially true, when the wavelength is longer than the thickness of the thin devices and the effective absorbing materials are not in abundance.[9] However, the electromagnetic properties of these checkerboard structures with extremely tiny connections have never been investigated in a realistic system. Shrinking the areas of the tiny connections between the squares results in a perfect self-complementary checkerboard pattern. However, with a low fabrication accuracy, the squares in the checkerboard pattern separate before the electromagnetic properties at the limit of the extremely small connections can be observed. Therefore, when the checkerboard pattern approaches the self-complementary pattern, the electromagnetic response abruptly changes.[10–13] To observe the electromagnetic response at the limit of the perfect self-complementary checkerboard pattern, during fabrication the extremely small connections must be maintained.

In this paper, we experimentally and numerically investigate the electromagnetic properties of a near perfect self-complementary checkerboard pattern with nano-sized connections, and show that the terahertz energy loss spontaneously becomes frequency independent. At the limit of a perfect self-complementary checkerboard pattern in a realistic system, the finite conductivity and randomness in the structure needs to be considered. Although metals can be treated as almost perfect electric conductors (PECs), because of their high conductivity at terahertz frequencies, the resistivity of the connecting points between the squares increases as the volume of the connection is reduced to nanometer-size. Here, we show for the first time that the PEC approximation spontaneously breaks down in a near perfect self-complementary checkerboard pattern, associated with broadening and maximization of the absorption spectrum. Under these conditions, the broadband absorption is still extremely concentrated at the nano-sized connection points. It is also revealed that a similar broadband energy loss results from the randomness in the connections between neighboring metallic squares. The scattering loss due to this randomness, contributes to the overall energy loss in these near perfect self-complementary checkerboard patterns.

## II.     Sample and measurement

To investigate the transition in the electromagnetic responses in the limit of extremely small connections, the checkerboard patterns were fabricated as shown in Fig. 1(b) by electron-beam lithography on a high-resistivity Si substrate of refractive index $n_{Si}$ = 3.42 and thickness $h_{Si}$ =

890 μm. The total dimensions of each pattern were 12 × 12 mm$^2$. The patterns were composed of gold on a titanium adhesion layer with thicknesses $h_{Au}$ = 450 nm and $h_{Ti}$ = 50 nm. Three checkerboard patterns with different square dimensions were fabricated while maintaining a constant period $p$ = 100 μm. The length of the square sides $d$ was controlled by varying the electron dose while nearly satisfying the self-complementarity condition: $d_0 = p/\sqrt{2} \sim 70.7$ μm. By increasing the electron dose, the exposed area on the resistive layer and the area of the deposited metal are both increased. Fig. 1 (c)-(e) show typical images for three samples S1, S2, and S3, obtained by scanning electron microscopy (SEM). In S1 and S3, the metallic squares are almost fully separated and connected, respectively. S2 most closely resembles a self-complementary checkerboard pattern.

The frequency-independent response of the self-complementary pattern is explained from Babinet's principle.[14,15] Babinet's principle requires a structure symmetric on both sides of the pattern. In Ref. 5, the checkerboard patterns on the substrate was covered by the substrate material to keep the symmetric structure. In this paper, the electromagnetic responses were measured without the covering materials. Although Babinet's principle is not strictly applicable to the patterns on the substrates owing to the asymmetric structure, the transmission and reflection coefficients approximately obey Babinet's principle as shown later. The transmission and reflection coefficients were measured by a terahertz time-domain spectroscopy system.[16] Photoconductive antennas were used as an emitter and detector and were excited by a femtosecond-pulse fiber laser. Using two parabolic mirrors, the emitted terahertz pulses were collimated, and the pulses transmitted through the samples were focused onto the detector antenna. The terahertz pulses were lineally polarized as shown in the schematics of Fig. 1(a). The beam diameter was approximately 8 mm on the sample. By applying Fourier transform to the time-domain waveforms, the amplitude and phase spectra were obtained.

### III. Results and discussion

**Nano-sized connections**

Figure 2(a) shows the real and imaginary parts of the complex transmission coefficients normalized by those of the Si substrate. Note that we assume a harmonic dependence $e^{-i\omega t}$ throughout this paper. The metallic squares in the sample S1 are separated, and so S1 is an array of metallic square chips. Because the currents on the small chips cannot reflect electromagnetic waves with a wavelength longer than the chips, the transmission coefficient for S1 approaches unity as the frequency decreases. In S1, the transmission coefficient decreases as the frequency increases and most electromagnetic waves around 0.85 THz are reflected by the resonance in the chip array. The metallic squares in S3 are connected to form a structure complementary to that of S1, making S3 an array of metallic holes. The transmission coefficient for S3 is also complementary to that of S1. This relation is known as Babinet's principle:[15,17,14]

$$\tilde{t}_o + \tilde{t}_c = 1 \text{ and } \tilde{r}_o + \tilde{r}_c = -1. \quad (1)$$

Here, $\tilde{t}_o$ ($\tilde{r}_o$) and $\tilde{t}_c$ ($\tilde{r}_c$) are the complex transmission (or reflection) coefficients for the original and complementary two-dimensional patterns, respectively. The transmission coefficients are normalized with the Fresnel transmission coefficient for an air-substrate interface, $\tilde{t}_{as} = 2/(n_{Si} + 1)$. Although the structure is asymmetric owing to the substrate, the transmission coefficients for two complementary patterns on the substrate roughly satisfy Eq. (1), as shown in Figs. 2(a) and 2(b). The sum of the transmission coefficients for the two complementary patterns on the substrate is roughly equal to the Fresnel transmission coefficient for the substrate without the patterns.

The transmission coefficients of S1 and S3 are reproduced well by the electromagnetic simulation, assuming they are made of PECs. The dashed lines in Fig. 2(a) show the complex transmission coefficients simulated by the finite-element method (using Ansys HFSS). In the simulation, the metallic parts of the checkerboard patterns were defined as PECs with a thickness of 450 nm and square sides of length $d = d_0 + \Delta d$, where $\Delta d = -100$ or $100$ nm for the separated or connected checkerboard patterns, respectively. As shown in Fig. 2(a), the PEC simulations reproduced the experimental results accurately. Although $\Delta d$ in the simulations is clearly larger than in the fabricated samples, the small structural discrepancy does not affect the transmission spectra. A PEC is a good approximation to metals in the terahertz region. The structural difference between samples S1 and S3 is approximately 100 nm, less than 0.1% of the wavelength (600 μm at 0.5 THz). The entire electromagnetic response fundamentally depends on whether the metallic squares are connected or not because of the high conductivity of Au at terahertz frequencies.

Since there is no energy absorption in the PEC, the sum of the energy transmission ($T$) and reflection ($R$) coefficients should be unity: $T + R = 1$. Combining this condition with Eq. (1), yields[5,17]

$$|\tilde{t}_i - 1/2| = 1/2, (i = o, c). \qquad (2)$$

Eq. (2) describes a circle of radius 1/2 centered on (1/2, 0) in the complex plane. The transmission coefficients of a 2-dimensional metallic pattern with zero energy loss are always located on this circle as the simulation results for PEC draw a circle in Fig. 2(b). Although the radius becomes slightly smaller owing to finite conductivity, the trajectories of the complex transmission coefficients for samples S1 and S3 are also on this circle, as shown in Fig. 2(b). The PEC is a good approximation of Au and there was small energy absorption in S1 and S3, though the patterns include nano-sized structures. The energy loss spectra, calculated by $1 -$ $T - R$, are plotted in Fig. 2(c) and indicated a small degree of absorption by S1 and S3, below the first-order diffraction frequency $f_c \sim c/(n_{Si}p) \sim 0.87$ THz. The circle in Fig. 2(b) corresponds to this regime. The energy loss caused by diffraction forces the circular arcs associated with S1 and S3 to shrink toward the center above $f_d$.

In contrast to S1 and S3, sample S2 exhibits a transmission spectrum that approaches the frequency-independent transmission coefficient of 1/2 in the self-complementary checkerboard pattern. The trajectory of the complex transmission coefficients for S2 forms a small ellipse in the complex plane [Fig. 2(b)]. This shape, smaller than the circle of radius 1/2, is associated with the energy loss indicated in Fig. 2(c). The complex transmission and reflection coefficients, of a film of impedance $z_f$ on the Si substrate, are described by $\tilde{t}_f = 2/(1 + n_{Si} + z_0/\tilde{z}_f)$ and $\tilde{r}_f = (1 - n_{Si} - z_0/\tilde{z}_f)/(1 + n_{Si} + z_0/\tilde{z}_f)$, respectively.[18] Here, $z_0 \sim 377\,\Omega$ is the vacuum impedance. Thus, the energy absorption by the thin film on the Si substrate is $A = 1 - |\tilde{t}_f|^2 n_{Si} - |\tilde{r}_f|^2 = |(4z_0/\tilde{z}_f)/(1 + n_{Si} + z_0/\tilde{z}_f)^2|$. This absorption reaches a maximum value $A = 0.226$ for $\tilde{z}_f \sim 85.3\,\Omega$ with $n_{Si} = 3.42$. Note that if the film is free-standing ($n_{Si} = 1$), the maximum value $A = 1/2$ is obtained for $\tilde{t}_f = -\tilde{r}_f = 1/2$ and $\tilde{z}_f = z_0/2$. The response of S2 in Fig. 2(c) shows that maximum broadband absorption is almost achieved due to the resistance of the nano-sized connections. Equation (1) can be written as $\tilde{t}_{z1} + \tilde{t}_{z2} = \tilde{t}_{as} = 2/(n_{Si} + 1)$ using the transmission coefficients, $\tilde{t}_{z1}$ and $\tilde{t}_{z2}$ for two complementary patterns with effective impedances $z_1$ and $z_2$ on the substrate. If two patterns are self-complementary, then ($\tilde{t}_{z1} = \tilde{t}_{z2}$), $\tilde{z}_1 = \tilde{z}_2 = z_0/(n_{Si} + 1) \sim 85.3\,\Omega$. The absorption is maximized in self-complementary patterns also on substrates.

The absorption of a thin metal film on a substrate can be maximized by tuning the film thickness and roughness.[18,19] However, we emphasize that the energy absorption is concentrated near the connection points in the near perfect self-complementary checkerboard patterns. Fig. 3 shows the full wave simulation (using, Ansys HFSS), of the transmission and energy absorption spectra for checkerboard patterns in which the Au squares are connected with nano-sized cuboids. The complex conductivity of Au, $\sigma_{Au}$, was governed by Drude dispersion in the simulation.[20] The absorption spectra [Fig. 3(d)] were obtained by integrating the Ohmic loss, calculated as the inner product of the electric-field and current distributions. The cuboid width and length were fixed at 5 nm. Over a range of 100 nm, the Au squares were tapered down toward the top of the small cuboids of height and width $h_c$, as displayed in the inset of Fig. 3(a). As $h_c$ decreased, the transmission peak at 0.84 THz also decreased, while the transmission below 0.6 THz increased and the absorption spectra increased, as shown in Fig. 3(c). Although the transmission spectra become flatter and the energy absorption increased with decreasing $h_c$, the height of the cuboid $h_c$, which is limited to being greater than 5 nm in our particular workstation, is not sufficiently thin to yield the flat energy absorption expected by the self-complementary patterns. However, when the cuboids are assumed to be metallic with conductivity $\tilde{\sigma}_{Au}/10$, the maximum broadband absorption was obtained, as shown in Fig. 3(d). This enhancement of the absorption stems from the increase in the resistivity of the cuboids and the resulting confinement of the electromagnetic waves within the cuboids.

Figure 4(a) shows the absorption coefficients calculated at 0.84 THz for each individual region in the above checkerboard patterns, as functions of the cuboid height $h_c$. The ohmic loss was integrated in each region. For $h_c$ values greater than 100 nm, the absorption is distributed in all metallic parts. As $h_c$ decreases, absorption increases in both the tapered region and cuboids, but decreases slightly in the other region. Although the cuboid volume, 5 nm × 5 nm × $h_c$, is extremely small compared to the other parts of the structure (a factor of $10^{-7}$ times the entire metallic square volume), more than 80% of the absorption is concentrated at the top of the cuboids and tapered parts when $h_c$ is small. Consequently, the currents become concentrated at the top of the cuboids as $h_c$ decreases from 50 to 5 nm, as shown in Figs. 4(b)-(d). When the metallic squares are fully separated [$h_c$ = 0, Fig. 4(e)], the current flow is obstructed and no electromagnetic energy is consumed. This is in contrast to the case with small finite $h_c$.

The resistivity at the top of the cuboids dominates the entire electromagnetic response of the near-self-complementary checkerboard patterns, as shown in the current distributions. A simple estimate of the complex impedance of a top cuboid is $\tilde{\sigma}_{Au}^{-1} \times length \times width$ which yields the complex cuboid impedance $(2.43 \times 10^{-8} - i1.88 \times 10^{-9})/h_c$ [Ω] and $(2.43 \times 10^{-8} - i3.75 \times 10^{-9})/h_c$ [Ω] at 0.50 and 1.0 THz, respectively. The Drude dispersion of the complex conductivity $\tilde{\sigma}_{Au}$ was used here.[20] There is almost no dispersion in the real part of the impedance, and the imaginary part is less than 1/5 of the real part in this so-called Haggen-Rubens frequency regime.[21] To obtain the required impedance of 85.3 Ω for maximum absorption, $h_c$ = 0.28 nm is required. This thickness is comparable to the atomic distance and is too thin to be fabricated as a continuous film. Assuming a conductivity of $\tilde{\sigma}_{Au}/10$, the cuboid impedance is 48.6 –$i$7.51 Ω for $h_c$ = 5 nm, which satisfies the condition of 85.3 Ω, and results in broadband absorption, shown in Fig. 3(d). In the actual samples, the surface roughness on the thin metal films can effectively decrease the conductivity $\tilde{\sigma}_{Au}$,[22] and the conductivity $\tilde{\sigma}_{Ti}$ of the adhesive Ti layer is less than $\tilde{\sigma}_{Au}/20$. Thus, the 85.3 Ω condition can be approached in the near-self-complementary checkerboard patterns, as sample S2 displays broadband energy loss. We also note the effect of the randomness in the connection cannot be ignored as discussed later.

We again emphasize that broadband and maximized absorption can be concentrated in a nano-sized area by using near-self-complementary checkerboard patterns. In Fig. 3, the transmission and absorption coefficients of a bowtie antenna array with nano-sized connections are compared with those of the checkerboard patterns. The absorption on the bowtie antenna array is also maximized but only near the resonant frequency [Fig. 3(c)]. It is worth also emphasizing that broadband absorption of the near-self-complementary checkerboard is confined to the nano-sized connections as showed in Fig. 4. In the previous report by Urade et al., the concentration of the broadband absorption in an area of 30 × 30 μm$^2$ of the checkerboard pattern with a unit cell of 150 × 150 μm$^2$ was achieved.[23] Fig. 4(a) indicates that further absorption concentration can be achieved by using a near-self-complementary checkerboard pattern. The finite conductivity leads to small absorption on the metal patches and tapered parts, except for the small connection parts, and the area of the absorption concentration is blurred. However, over 80% of the broadband absorption can still be concentrated in the deep subwavelength area, less than 200 × 200 nm$^2$ ($\lambda$ = 600 μm at 0.5 THz) within the unit cell of 100 × 100 μm$^2$. If we define a concentration factor as the ratio of unit cell area to absorption area, the concentration factor $2.5 \times 10^5$ in this work, is four orders larger than that of 25, reported in previous work.[23] The near-self-complementary checkerboard pattern works as a broadband interface connecting two extremely different scales. The absorption coefficient in any 2-dimentional system including the self-complementary structure is limited to be less than 0.5. Although the additional 3-dimensional structures are generally required to achieve the absorption efficiency more than 0.5, the near-self-complementary design can be a base structure for such 3-dimentional system because of their broad-bandwidth and energy concentration.

**Randomness in connections**

We show that randomness in the connections also produces broadband energy loss. The energy loss in sample S2 includes the scattering loss caused by randomness arising from the coexistence of connected and unconnected Au squares in the near-self-complementary checkerboard patterns. It is shown here that the randomness of the connections can effectively maximize the broadband energy loss without absorption. the metallic checkerboard patterns were defined to be of finite-size $10 \times 10$ mm$^2$ and the complex transmission coefficient and scattering loss were calculated using the finite-difference time-domain method (Fujitsu, Poynting for optics). The metallic squares were composed of a PEC with zero thickness, a period of 300 μm, and side length $300/\sqrt{2} - 4 \sim 208$ μm. Connections between neighboring squares were randomly generated from the small PEC chips with probability $a$. In the limit $a = 0$, all metallic squares were fully separated, whereas for $a = 1$, they are all connected with small squares of area $8 \times 8$ μm$^2$, as shown in Fig. 5. The checkerboard pattern was irradiated by a Gaussian beam with a standard deviation of 4 mm. The transmitted and reflected electric fields were calculated for various $a$ values on a plane, of area $4 \times 4$ mm$^2$ at a distance 2.9 mm away from the checkerboard pattern. The time-domain waveforms were averaged over the observation plane. This operation corresponds to the focusing of coherent transmitted waves onto a single-pixel detector. The averaged time-domain waveforms were Fourier-transformed and normalized by those without the checkerboard patterns, to calculate the complex transmission and reflection coefficients. The energy loss was calculated from these coefficients and is defined as $1 - |\tilde{t}|^2 - |\tilde{r}|^2$. Because the patterns were defined by the PEC, the energy loss is only caused by scattering.

The transmission coefficients corresponding to probabilities $a$ and $1 - a$ are mutually complementary, as shown in Fig. 5. With increasing connection probability $a$, the energy loss increases and becomes maximized for $a = 0.5$. For intermediate $a$ values, the pattern includes both connected and separated checkerboard patterns. The Babinet's principle indicates that scattering fields from two mutually complementary patterns have the same amplitude, but are in anti-phase. Therefore, the scattering fields from the connected and separated checkerboard patterns are cancelled out on the detector. A checkerboard randomly connected with $a = 0.5$ has, on average, the same structure as its complementary checkerboard, reflecting self-complementary. As a result, the transmission coefficient and energy loss exhibits a broadband flat spectrum of height 0.5 for $a = 0.5$, in which the number of connected squares equal that of the separated. The scattering loss can also result in a broadband spectrum and maximized energy loss, as well as absorption. Tremain et al., also reported that the number of connected squares significantly governs the transmission, reflection, and energy loss by introducing rotational disorder into the checkerboard patterns.[24] They also found that approximately 50% of connections exhibit broadband dispersion-less energy loss. Scattering loss was included in our experimental results shown in Fig. 2. Here, it should be noted that energy loss due to scattering and absorption cannot be separated experimentally in terahertz time-domain spectroscopy.

A similar effect caused by randomness can be seen in the optical response of metallic films. Metallic films at the percolation threshold exhibit broadband flat transmission, reflection, and absorption spectra.[25] At the percolation threshold, the films form fractal shapes consisting of infinitely long clusters,[26] which display frequency-independent behavior because of their scale invariance.[27] At the critical probability $a = 0.5$, the patterns form fractal clusters despite being finite. Broadband dispersion-less responses can be also understood in terms of the structural self-similarity of the percolation films.[13]

## IV.    Conclusions

Because of the very high conductivity of metals in the terahertz region, extremely small structural changes can induce a transition in the response of checkerboard systems. This transition, however, is not discontinuous. The effects of nano-sized connections and the finite conductivity of the metal connecting the squares were investigated, both experimentally and numerically in the terahertz frequency range. Optimal nano-sized connections produce maximum broadband absorption. The broadband absorption is strongly confined to the deep subwavelength area, less than $200 \times 200$ nm$^2$, in the near-self-complementary checkerboard patterns with finite conductivity. These are promising for applications of terahertz devices, such as bolometers, sensors, and modulators.[6,7,28] However, the effects of randomness in the connections between neighboring squares in the checkerboard patterns are not negligible. These are important in inducing apparent energy loss in the transmission spectrum obtained of the zeroth-order scattered waves. It was revealed that, as well as the finite conductivity of metals, scattering loss due to randomness effectively achieves the zeroth order transmission coefficient $\tilde{t} = 1/2$, a reflection coefficient of $\tilde{r} = -1/2$, and a broadband energy loss of $A = 1/2$.

An interesting future application of the self-complementary pattern is the coherent control of absorption and transmission. When a film with $\tilde{t} = 1/2$ and $\tilde{r} = -1/2$ is irradiated from both sides with coherent electromagnetic waves, absorption can be controlled via the phases of the incident waves. If the two incident beams are in phase, the transmitted and reflected waves interfere destructively, so that all the energy is absorbed into the film. If instead the two incident beams are in antiphase, then there is no absorption. Perfect absorption and signal processing have already been demonstrated by using a metasurface,[29] a multi-layer graphene film,[30] and a checkerboard pattern with resistive connections.[23] Checkerboards with nano-sized connections or randomness are also promising candidates for achieving broadband coherent control of the transmission, reflection, and confined absorption or scattering loss.

**Acknowledgments**

The authors thank Dr. Y. Urade for helpful discussions. This work was supported by Japan Society for the Promotion of Science (JSPS) KAKENHI (16H06025, 16H03886, 25286063) from the Ministry of Education, Culture, Sports, Science, and Technology (MEXT) KAKENHI (22109003).

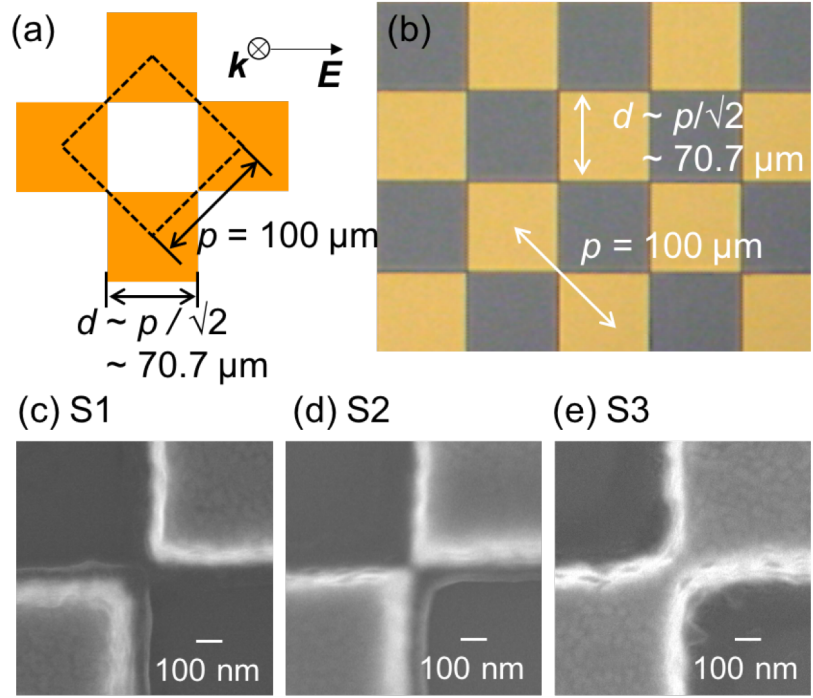

FIG. 1. (a) Schematic and (b) microscope image of a typical checkerboard pattern fabricated by EB-lithography on a Si substrate. (c)-(e) Typical SEM images of the connections between Au squares produced by three different electron doses: samples (c) S1, (d) S2, and (e) S3.

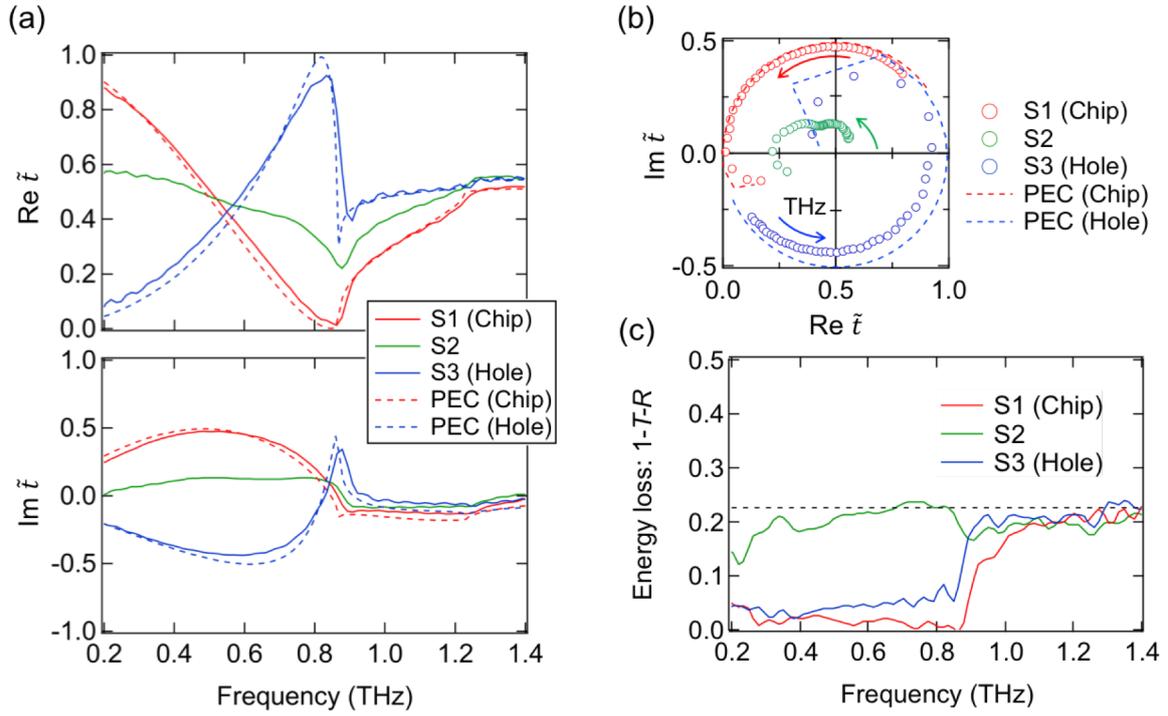

FIG. 2. (a) The real and imaginary parts of the complex amplitude transmission coefficients of the checkerboard patterns. The experimental and simulation results obtained for PECs are indicated by the solid and dashed lines, respectively. The coefficients are normalized with those of the substrate. (b) Cole-Cole plots of the complex transmission coefficients in the frequency range from 0.2 to 0.9 THz. The circles and dashed lines indicate the experimental and simulation results, respectively. The arrows indicate the direction of increasing frequency. (c) The energy-loss spectra, calculated by $1 - T - R$. The dashed line indicates the maximum absorption level 0.226 for $n_{Si}= 3.42$.

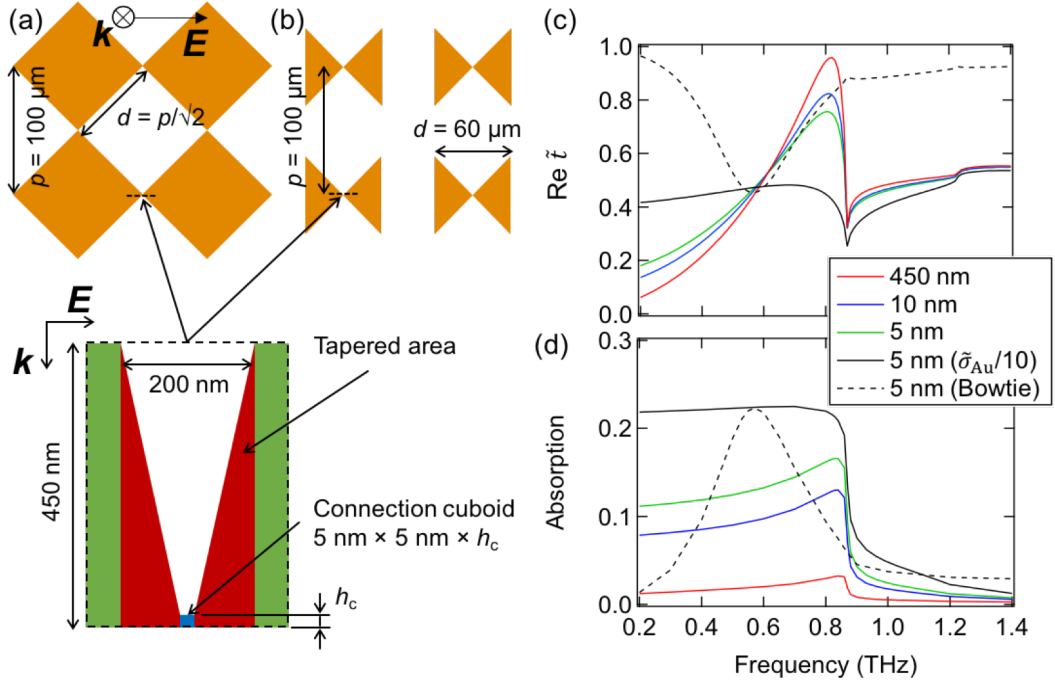

FIG. 3. Schematics of the simulation models for (a) checkerboard patterns and (b) bow-tie arrays with nano-sized connections. The inset shows the cross-section of the region between the squares and triangles, which are connected by tapered regions and a nano-sized cuboid. (c) Real part of the complex transmission coefficients and (d) absorption spectra. The red, blue, and green lines denote the simulation results, assuming $h_c$ = 450, 10, and 5 nm, respectively. The black lines indicate the simulation results with nano-sized connections with a conductivity of 1/10 that of Au and $h_c$ = 5 nm. The black dashed lines indicate the simulation results for the bow-tie arrays with $h_c$ = 5 nm.

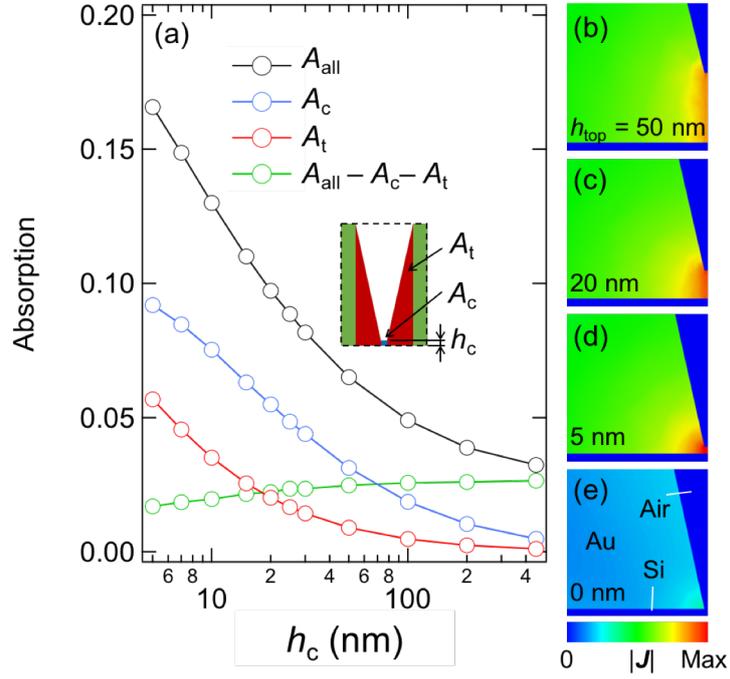

FIG. 4. The calculated absorption coefficients as a function of $h_c$ at 0.84 THz, for each structural part. The joule loss was calculated for the entire structure (black, $A_{all}$), nano-sized cuboid (blue, $A_c$), tapered parts (red, $A_t$), and all remaining parts (green, $A_{all} - A_c - A_t$). The inset shows the cross section of the connections, including the tapered parts and cuboid. The current distributions were calculated in the cross section for (b-e) $h_c$ = 50, 20, 5, and 0 nm. For $h_c$ = 0 nm, there was no connection between the squares.

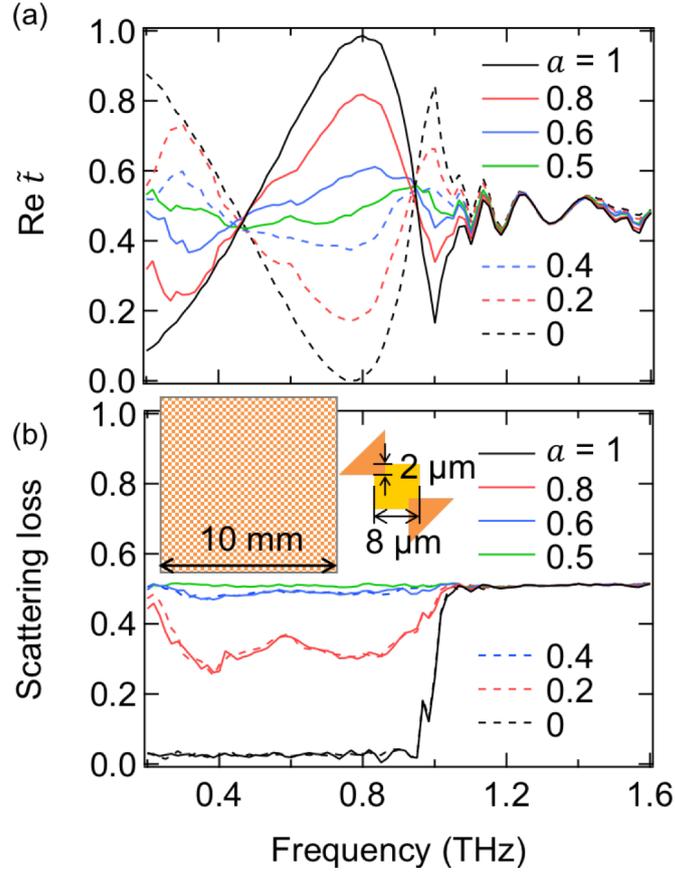

FIG. 5. (a) The real part of the complex transmission spectra and (b) the energy-loss spectra calculated for checkerboard patterns with a finite area (10 × 10 mm$^2$) assuming random connections. The metallic squares (208 × 208 μm$^2$) are aligned with a period of 300 μm. The tops of the metal squares are connected via small squares, with a probability $a$. The small chips (8 × 8 μm$^2$) were placed randomly between the tops of the larger squares, as shown in the inset. Both the large and small squares are PECs with zero thickness.